\documentclass[runningheads]{llncs}
\usepackage[T1]{fontenc}
\usepackage{graphicx,verbatim}
\usepackage{cite}
\usepackage{booktabs}
\usepackage{xcolor}

\begin{document}
\title{A Non-contrast Head CT Foundation Model for Comprehensive Neuro-Trauma Triage}
\titlerunning{Comprehensive Neuro-Trauma Triage}
%
\author{Youngjin Yoo\inst{1} \and
Bogdan Georgescu\inst{1} \and
Yanbo Zhang\inst{1} \and
Sasa Grbic\inst{1} \and
Han Liu\inst{1} \and
Gabriela D. Aldea\inst{1,8,9} \and
Thomas J. Re\inst{1} \and
Jyotipriya Das\inst{1} \and
Poikavila Ullaskrishnan\inst{1} \and
Eva Eibenberger\inst{2} \and
Andrei Chekkoury\inst{2} \and
Uttam K. Bodanapally\inst{4} \and
Savvas Nicolaou\inst{5} \and
Pina C. Sanelli\inst{6} \and
Thomas J. Schroeppel\inst{7} \and
Yvonne W. Lui\inst{3} \and
Eli Gibson\inst{1}}
\authorrunning{Yoo et al.}

%
\institute{Digital Technology and Innovation, Siemens Healthineers, Princeton, NJ USA \and
Department of Computed Tomography, Siemens Healthineers, Forchheim, Germany \and
Department of Radiology, New York University, New York, NY USA \and
Department of Radiology, University of Maryland Medical Center, Baltimore, MD USA \and
Department of Radiology, Vancouver General Hospital, Vancouver, BC Canada \and
Department of Radiology, Northwell Health, New York, NY USA \and
Department of Surgery, UCHealth Memorial Hospital, Colorado Springs, CO USA \and
Foundational Technologies, Siemens SRL, Braşov, Romania \and
Automation and Information Technology Department, Transilvania University of Braşov, Braşov, Romania
}

%
\maketitle              
\begin{abstract}
Recent advancements in AI and medical imaging offer transformative potential in emergency head CT interpretation for reducing assessment times and improving accuracy in the face of an increasing request of such scans and a global shortage in radiologists. This study introduces a 3D foundation model for detecting diverse neuro-trauma findings with high accuracy and efficiency. Using large language models (LLMs) for automatic labeling, we generated comprehensive multi-label annotations for critical conditions. Our approach involved pretraining neural networks for hemorrhage subtype segmentation and brain anatomy parcellation, which were integrated into a pretrained comprehensive neuro-trauma detection network through multimodal fine-tuning. Performance evaluation against expert annotations and comparison with CT-CLIP demonstrated strong triage accuracy across major neuro-trauma findings, such as hemorrhage and midline shift, as well as less frequent critical conditions such as cerebral edema and arterial hyperdensity. The integration of neuro-specific features significantly enhanced diagnostic capabilities, achieving an average AUC of 0.861 for 16 neuro-trauma conditions. This work advances foundation models in medical imaging, serving as a benchmark for future AI-assisted neuro-trauma diagnostics in emergency radiology.

\keywords{Foundation Model \and Neuro-Trauma Detection \and Head CT.}
\end{abstract}

\section{Introduction}
\label{sec:introduction}
Head computed tomography (CT) is an essential diagnostic tool of emergency medicine, particularly for assessing acute neurological symptoms and head trauma~\cite{lolli2016mdct,sharp2017computed}. Its utilization is on the rise~\cite{yun2018utilization} while the availability of trained radiologists qualified to interpret its results is facing a worldwide shortage~\cite{yaghmai2024editorial,parag2022shortage}. AI-assisted interpretation of emergent head CT could help address this situation by increasing the efficiency and accuracy of available qualified radiologists and of less specialized clinicians in interpreting such imaging~\cite{vimalesvaran2024assessing,savage2024prospective}. AI-driven approaches have evolved from supervised methods~\cite{krizhevsky2012imagenet,simonyan2014very} to self-supervised and semi-supervised approaches~\cite{ronneberger2015u,he2016deep}, reducing dependence on extensive annotations. Recently foundation models trained on large datasets have shown remarkable success across domains~\cite{bommasani2021opportunities} such as MedViT~\cite{manzari2023medvit} and MIMIC-CXR~\cite{johnson2019mimic}. Notably, CT-CLIP~\cite{hamamci2024foundation} has enabled supervised-level zero-shot detection of chest abnormalities and FM-CT~\cite{zhu20253d}, a newly introduced head CT foundation model for detecting various neuro conditions such as hemorrhages, tumors and other abnormalities, illustrating the transformative potential in medical imaging.

Despite this progress, applying foundation models to head CT remains challenging due to anatomical complexity and the broad spectrum of neuro-trauma conditions. Rapid and accurate diagnosis is critical in emergency settings~\cite{rincon2016imaging}, yet traditional interpretation is time-consuming and prone to variability~\cite{parag2022interpretation,erly2002radiology}. Existing foundation models may underperform in neuro-trauma detection due to domain-specific limitations. To address this, we developed a 3D foundation model specialized for head CT, trained on a large multi-site dataset. This model enables accurate, efficient few-shot detection of neuro-trauma conditions, potentially enhancing trauma triage and improving patient outcomes.

Our contributions include setting benchmark performance for neuro-trauma detection, demonstrating robust generalization across common and rare critical findings, and emphasizing the importance of domain-specific pretraining~\cite{yoo2023importance,zhu20253d}. By integrating neuro-specific pathological and anatomical features, we highlight the advantages of specialized foundation models over CT-CLIP and the broader coverage of neuro-trauma findings compared to FM-CT.

\section{The comprehensive neuro-trauma detection foundation model}
\label{sec:methods}

To develop a head CT foundation model, a neuro-radiologist curated a comprehensive list of neuro-trauma findings requiring urgent clinical attention~~\cite{wintermark2015imaging,zimmermann2019emergency}, including Hemorrhage, Infarct, Mass Lesion, Mass Effect, Hydrocephalus, Midline Shift, Skull Fracture, Cerebral hemorrhagic Contusion, Diffuse Cerebral Edema, Microhemorrhage, Diffuse Axonal Injury, Generalized Cerebral Edema, Pneumocephalus, Brain Herniation, Arterial Hyper-density and Venous Sinus Hyper-density. Using this list, we automatically generated labels for a large-scale dataset through an Large Language Models (LLMs) pipeline. We independently pretrained two task-specific vision networks and integrated these pretrained networks into a foundation model via multimodal finetuning with LLM-generated labels. Instead of training directly on image-report pairs~\cite{hamamci2024foundation}, we utilized image-LLM multi-label pairs to streamline task-specific pretraining and directly integrate anatomical and pathological features.

\subsection{Automatic comprehensive labeling}
\label{subsec:llm_labeling}
Leveraging recent advancements of LLMs in generating medical content~\cite{bicknell2024chatgpt}, we automatically generated multi-labels for neuro-trauma findings for each radiological report in our dataset. The labels were generated using a private GPT4-o model on our private network. To efficiently label a large-scale dataset, we investigated prompts that requested 16 multiple labels, minimizing the need for repetitive label processing. The optimized prompt was ``Given this radiology report, extract POS or NEG value for these concepts \{0\}. POS means the concept is present in patient as per report. NEG means not. Return as json format with keys being the name of the concepts and value being either POS or NEG. Report:\{{\it{report content}}\}'', and for \{0\} we provided the entire 16 neuro-trauma findings as a list. Presenting labeling examples to the LLM before labeling the entire dataset improved labeling accuracy.

\subsection{Task-specific pretraining}
\label{subsec:task_specific_pretraining}
We pretrained two networks independently, each performing a specific task: brain bleeding (hemorrhage) subtype segmentation and brain anatomy parcellation.

The hemorrhage subtype segmentation network is based on a 3D Dense U-Net architecture~\cite{huang2017densely}, specifically designed for classifying five hemorrhage subtypes: intraparenchymal, subarachnoid, intraventricular, subdural, and epidural hemorrhages~\cite{gibson2022artificial}. We adapted the specialized network architecture described in~\cite{gibson2022artificial} for this task. In this work, the network additionally incorporates Squeeze-and-Excitation (SE) blocks~\cite{jie2018squeeze} to enhance feature recalibration throughout the architecture. SE blocks are strategically placed before and after each DenseBlock to adaptively recalibrate feature responses.

The brain parcellation network is designed for segmenting 15 brain structures: left/right hemispheres, supratentorial/infratentorial regions, frontal lobe, parietal lobe, occipital lobe, temporal lobe, cerebellum, basal ganglia, medulla oblongata, pons, midbrain, falx and ventricles. The network employs a U-Net architecture with 15 stages, using ReLU activations, batch normalization, strided convolutions for downsampling and transposed convolutions for upsampling. The model employs a multi-head architecture with three output layers: one dedicated to segmenting left-side hemispheres, another for supratentorial-infratentorial regions, and a third handling the remaining brain structures.

\subsection{Building a foundation model with multimodal finetuning}
\label{subsec:finetuning}
\begin{figure}
\centering
\includegraphics[width=0.9\textwidth]{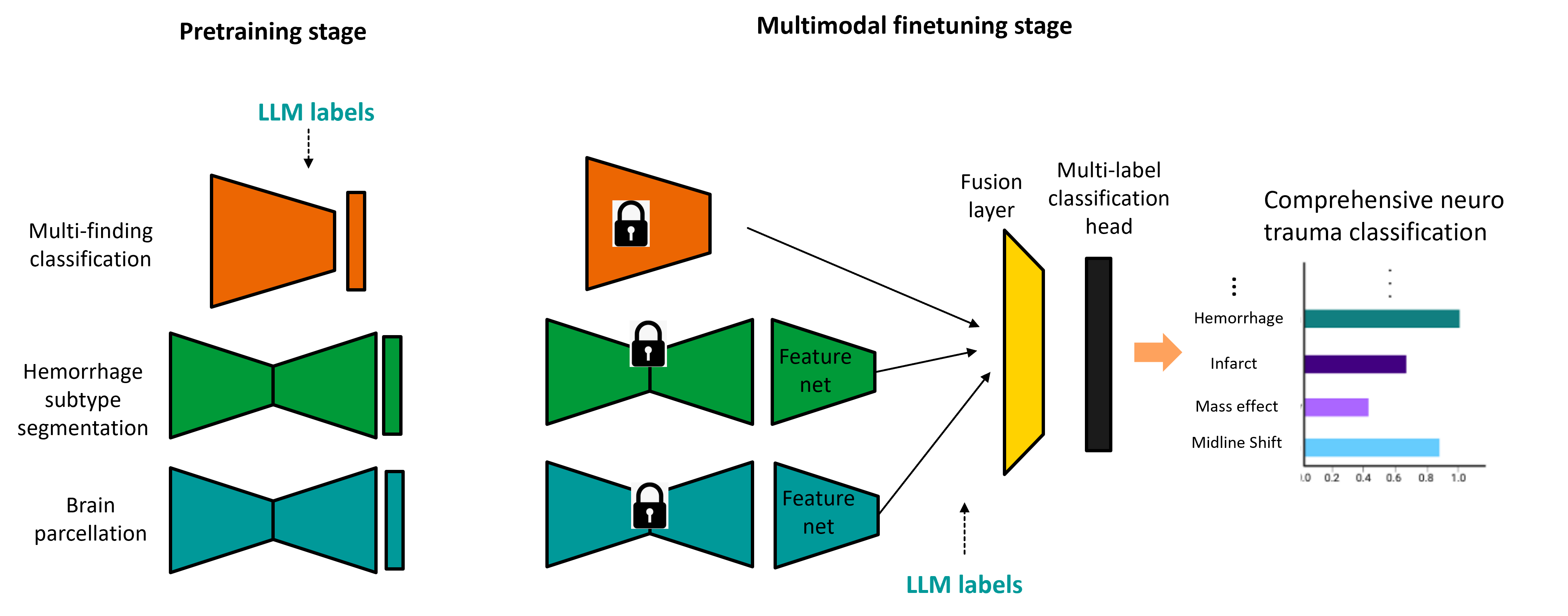}
\caption{Overview of the head CT foundation model for neuro-trauma triage.} \label{fig:model_overview}
\end{figure}

Our foundation model is based on the 3D densely connected network specifically designed for brain hemorrhage classification~\cite{gibson2022artificial}, which we trained to perform comprehensive neuro-trauma detection tasks. We refer to it as the Comprehensive Neuro Trauma Detection Network (CNTD-Net). To learn a wider variety of imaging features for handling heterogeneous pathologies beyond brain hemorrhage, we expanded the network capacity of the brain hemorrhage classification network by increasing the number of layers and feature channels, resulting in the DeepCNTD-Net. Specifically, we increased the convolutional channel growth rate of the 3D DenseBlock from 5 to 8, the initial number of feature maps from 16 to 64, the total number of 3D dense layers from 15 to 20, and the final feature vector dimension from 1638 to 4032. We pretrained both CNTD-Net and DeepCNTD-Net by performing the comprehensive neuro-trauma detection task using the LLM-generated multi-labels. To integrate the task-specific pretrained networks, we employed an encoder that extracts features with dimensions similar to those of DeepCNTD-Net by collapsing the segmentation features. The features from DeepCNTD-Net, the hemorrhage subtype segmentation network, and the brain parcellation network were then fused using linear layers, which were subsequently used for multi-label classification. The feature fusion process was performed after freezing the pretrained networks. Pre-training and fine-tuning were performed using the Adam optimizer \cite{Kingma2014AdamAM} to adapt the learning rates for each parameter. Additionally, binary cross-entropy with logits loss was employed, incorporating adjusted class weights to effectively address the class imbalances in the dataset. The overall procedure is illustrated in Fig.~\ref{fig:model_overview}.


\vspace{-20pt}

\section{Results and Discussion}

\subsection{Dataset and Preprocessing}

\begin{figure}
\centering
\includegraphics[width=0.8\textwidth]{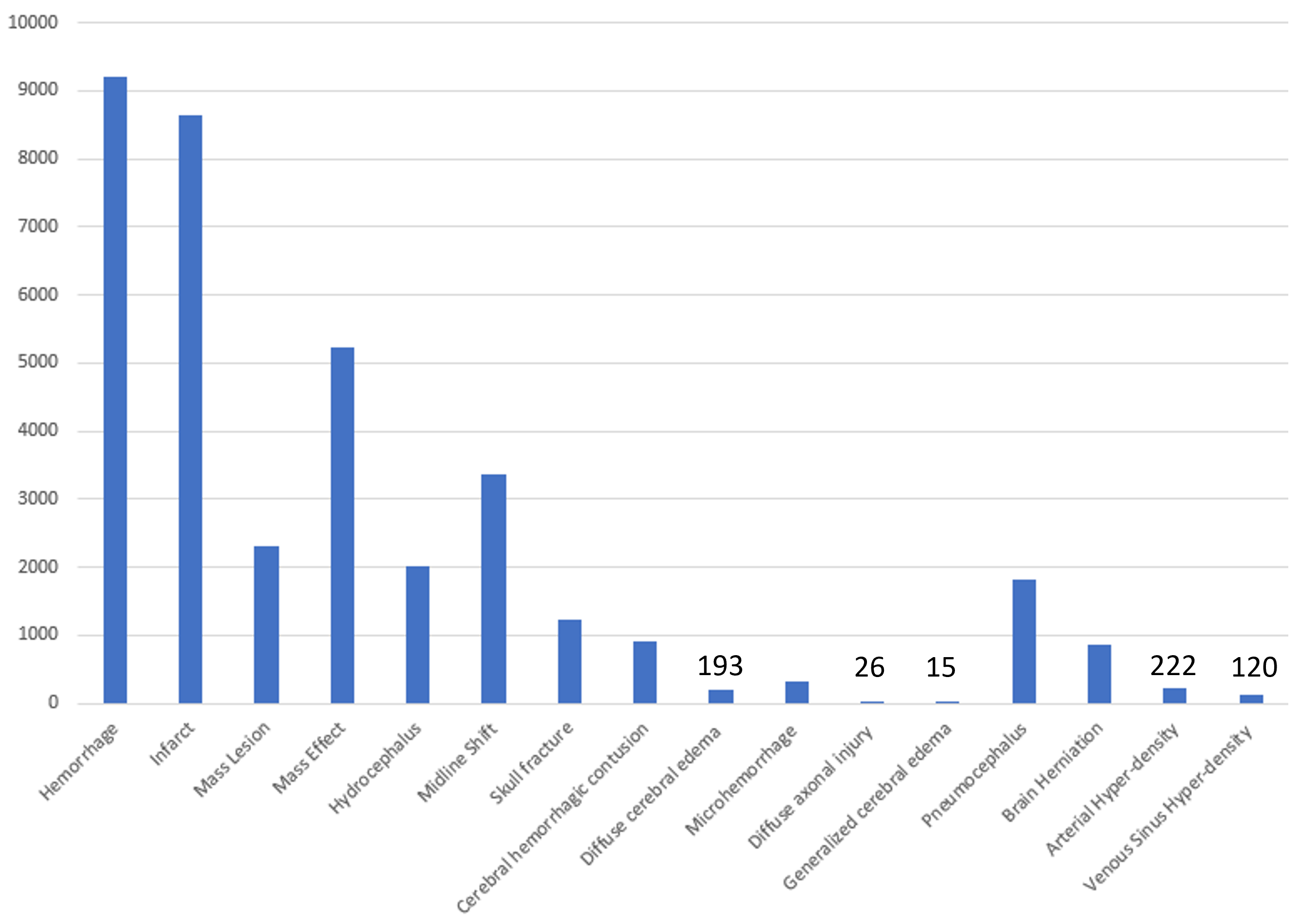}
\caption{A comprehensive list of neuro-trauma findings that encompass the critical conditions requiring immediate clinical attention in trauma emergency centers curated by a neuro-radiologist and their prevalence in our dataset.} \label{fig:prevalence}
\end{figure}

Anonymized non-contrast CT (NCCT) head volumes were retrospectively collected from nine centers across the U.S., Canada, China, and India, with ethics committee approvals waiving informed consent. Data were sourced from pre-established cohorts or retrospective selections. NCCT volumes were acquired using Siemens, GE, and Toshiba scanners. Exclusion criteria included age under 18 or absence of axial reconstruction. A total of 29,395 NCCT studies met inclusion criteria: (a) 26,514 studies were used for model development—23,592 for training and 2,922 for system optimization (architecture selection, parameter tuning, and classifier calibration); (b) 2,881 studies were reserved for independent performance evaluation. Patient-level random splitting was performed before development to prevent data contamination. The prevalence of trauma findings, based on LLM labels, is shown in Fig.~\ref{fig:prevalence}. Preprocessing involved automatic alignment of axial NCCT volumes to a standard reference frame, resampling to a 1-mm in-plane and 4-mm out-of-plane resolution, and normalization using Hounsfield Unit (HU) windows: 0–80 HU (bleeding), -20–180 HU (brain), and -800–2000 HU (bone), scaled to 0–1. To enhance robustness against translation and scanner noise, data augmentation included random in-plane translation (±10 mm), in-plane flipping (50\% probability), random CT windowing noise (±10 HU), and random image noise (0.01 STD).


\subsection{LLM labeling accuracy against manual expert label}
\label{subsec:LLM_accuracy}
We evaluated the accuracy of our LLM pipeline labeling against expert-generated manual labels for six major neuro-trauma findings in head CT scans (Table~\ref{tab:llm_acc_all}). These six major labels were previously generated by expert users. LLMs achieved 92–99\% accuracy, except for ischemia / infarction at 79\%, probably due to the lower sensitivity of NCCT causing ambiguities in the report. This highlights the LLMs’ robustness in detecting diverse pathologies. Table~\ref{tab:llm_acc_all} compares LLM labels with neuro-radiologist annotations for 200 randomly selected cases for all 16 findings. The LLMs excelled in identifying microhemorrhages, diffuse axonal injuries, and venous sinus hyperdensity (accuracy: 1.0) and performed well on diffuse/generalized cerebral edema and skull fractures (0.99). These results demonstrate the LLMs’ effectiveness in accurately labeling complex head CT findings, supporting automated trauma triage.


\begin{table}
\centering
\footnotesize 
\caption{LLM labeling accuracy compared to expert labels. Accuracy is reported for six major findings across the entire dataset (values in parentheses) and for all 16 findings in 200 randomly selected cases.}\label{tab:llm_acc_all}
\begin{tabular}{|l|c|l|c|}
\hline
Finding &  Accuracy & Finding &  Accuracy \\
\hline
Hemorrhage & 0.95 (0.920)  & Diffuse cerebral edema & 0.99 \\
Ischemia/Infarct & 0.80 (0.786) & Microhemorrhage & 1.0 \\
Mass Lesion & 0.95 (0.921) & Diffuse axonal injury & 1.0 \\
Mass Effect & 0.94 (0.958) & Generalized cerebral edema & 1.0 \\
Hydrocephalus & 0.95 (0.950) & Pneumocephalus & 0.96 \\
Midline Shift & 0.99 (0.987) & Brain herniation & 0.98 \\
Skull fracture & 0.99 & Arterial Hyper-density & 0.99 \\
Cerebral hemorrhagic contusion & 0.98 & Venous Sinus Hyper-density & 1.0 \\
\hline
\end{tabular}
\end{table}


\subsection{Comprehensive neuro-trauma detection performance}
\label{subsec:det_perf_results}
Table~\ref{tab:det_perf} presents ablation study results, assessing the impact of different model components. The baseline CNTD-Net, designed for six major findings, achieved an AUC of 0.768. Expanding it to DeepCNTD-Net significantly improved performance (AUC: 0.858). Adding brain hemorrhage segmentation features (hemSegFeat) further increased AUC to 0.873, underscoring their importance. Incorporating brain anatomy features (brainAnatFeat) provided a slight boost to 0.875. For detecting all 16 findings, DeepCNTD-Net reached an AUC of 0.849, improving to 0.859 with hemSegFeat and peaking at 0.861 with brainAnatFeat. Table~\ref{tab:det_perf_comparison} compares DeepCNTD-Net with CT-CLIP, which, despite being trained on the same LLM-generated labels and fine-tuned with CT-LiPro [6], achieved lower AUCs (0.822 for six major findings, 0.835 for all 16). In contrast, DeepCNTD-Net, leveraging hemSegFeat and brainAnatFeat, achieved superior scores of 0.875 and 0.861, highlighting its advantage in neuro-trauma detection through specialized feature integration.

\begin{table}
\centering
\footnotesize 
\caption{Average detection performance (AUC) results of ablation study for the six major neuro-trauma findings and all 16 findings. {\bf{hemSegFeat}}: brain hemorrahge segmentation features; {\bf{brainAnatFeat}}: brain anatomy segmentation features.}\label{tab:det_perf}
\begin{tabular}{|l|c|}
\hline
Model &  Average AUC\\
\hline
CNTD-Net (6 major findings) &  $0.768 \pm 0.064$ \\
DeepCNTD-Net (6 major findings) & $0.858 \pm 0.066$ \\
DeepCNTD-Net + hemSegFeat (6 major findings) & $0.873 \pm 0.068$ \\
DeepCNTD-Net + hemSegFeat + brainAnatFeat (6 major findings) & $0.875 \pm 0.065$ \\
DeepCNTD-Net (all 16 findings) & $0.849 \pm 0.090$ \\
DeepCNTD-Net + hemSegFeat (all 16 findings) & $0.859 \pm 0.085$ \\
DeepCNTD-Net + hemSegFeat + brainAnatFeat (all 16 findings) & $0.861 \pm 0.081$ \\
\hline
\end{tabular}
\end{table}

\begin{table}
\centering
\footnotesize 
\caption{Average detection performance of CT-CLIP and DeepCNTD-Net with hemSegFeat and brainAnatFeat.}\label{tab:det_perf_comparison}
\begin{tabular}{|l|c|}
\hline
Model &  Average AUC\\
\hline
CT-CLIP (6 major findings) & $0.822 \pm 0.081$\\
DeepCNTD-Net (6 major findings) & $0.875 \pm 0.065$ \\
CT-CLIP (all 16 findings) & $0.835 \pm 0.083$\\
DeepCNTD-Net (all 16 findings) & $0.861 \pm 0.081$ \\
\hline
\end{tabular}
\end{table}

\begin{figure}
\centering
\includegraphics[width=\textwidth]{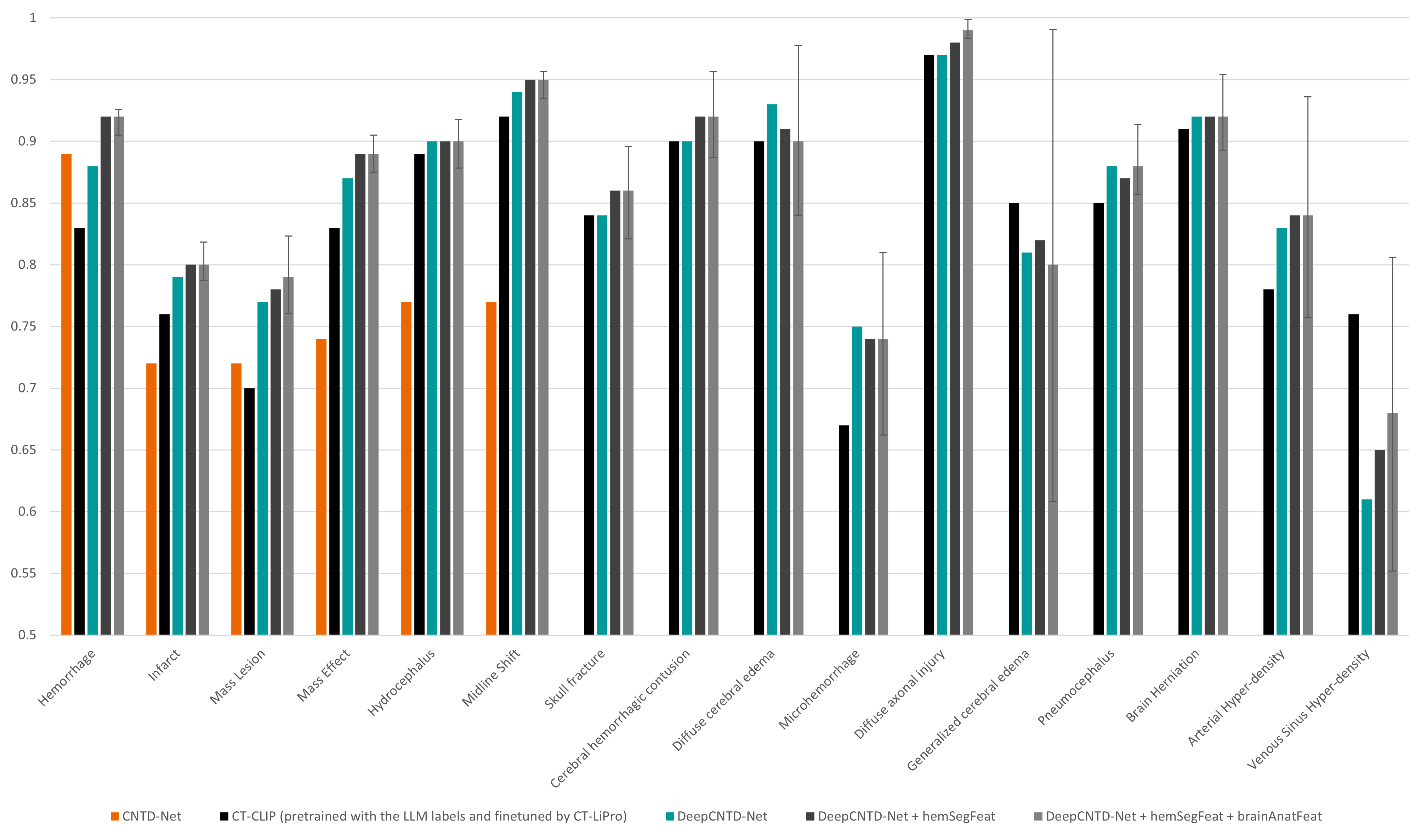}
\caption{Individual detection performance (AUC) for six major critical findings and all 16 neuro-trauma conditions. CNTD-Net was evaluated exclusively for the six major findings.} \label{fig:det_perf_comp}
\end{figure}

\begin{figure}
\centering
\includegraphics[width=0.90\textwidth]{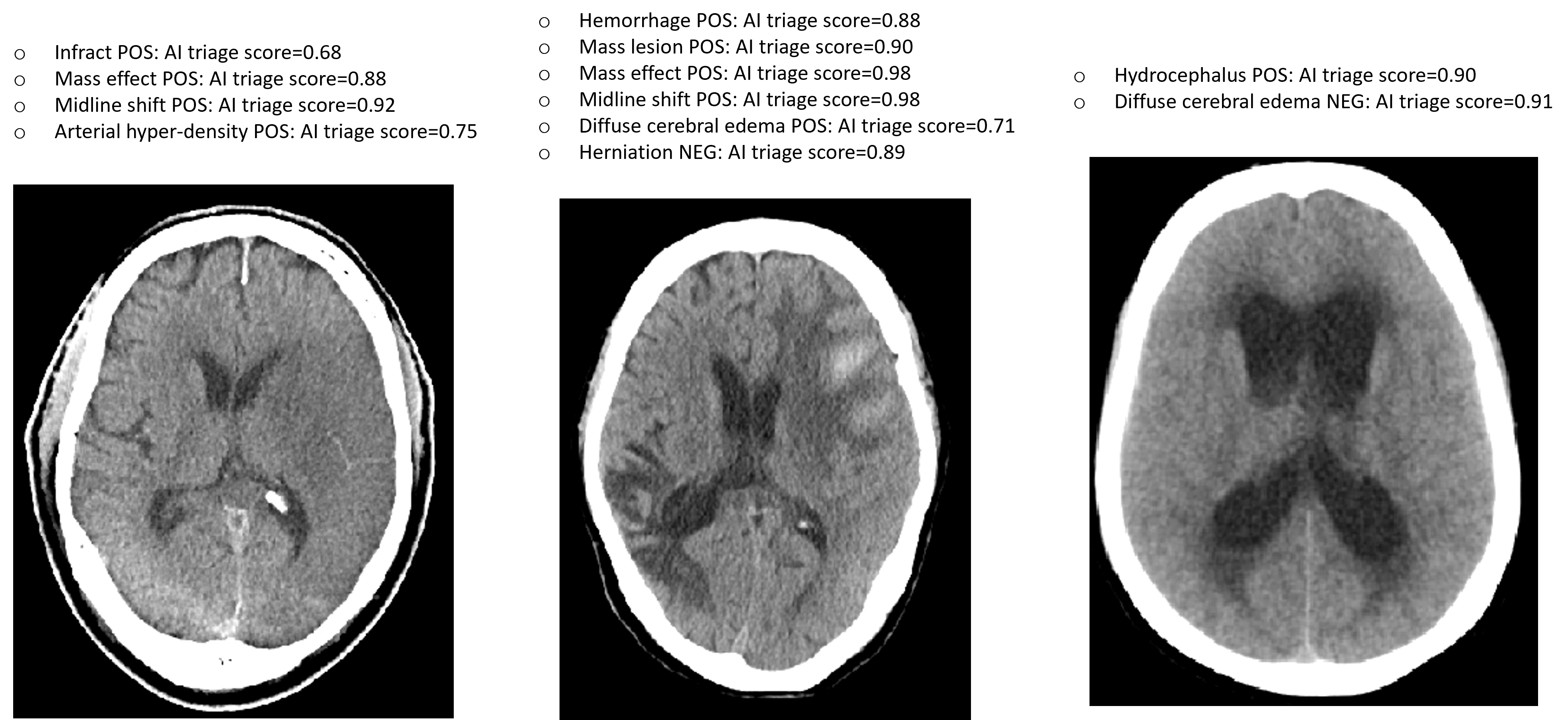}
\caption{Examples of comprehensive neuro-trauma triage. "POS" indicates a positive case, while "NEG" denotes a negative case, both based on the LLM-gerated multi-labels, and AI triage score is between 0.0 and 1.0.} \label{fig:det_examples}
\end{figure}

We analyzed individual AUC performance to identify key factors driving detection improvements (Fig.~\ref{fig:det_perf_comp}). Models incorporating segmentation and anatomical features consistently outperformed baselines across both major and rare findings, reinforcing the value of multimodal integration. DeepCNTD-Net, particularly with hemSegFeat and brainAnatFeat, delivered superior results, achieving an AUC of 0.92 for hemorrhage detection versus CT-CLIP’s 0.83 and FM-CT's range of 0.835-0.929 (reported in~\cite{zhu20253d}). For midline shift, DeepCNTD-Net variants reached up to 0.95, surpassing CT-CLIP’s 0.92. The enhanced models also improved mass effect detection (AUC: 0.89), while CT-CLIP excelled in generalized cerebral edema, suggesting complementary strengths that could be leveraged through model fusion. To assess generalizability, we evaluated DeepCNTD-Net on the CQ500 dataset~\cite{chilamkurthy2018deep}, where it maintained strong performance in hemorrhage (AUC: 0.920) and midline shift (AUC: 0.965), outperforming FM-CT~\cite{zhu20253d} (hemorrhage AUC: 0.776-0.850, midline shift AUC: 0.780), but exhibited lower accuracy for mass effect (AUC: 0.840) against FM-CT (AUC: 0.90). Additionally, FM-CT outperformed in detecting edema (AUC: 0.827-0.923) and hydrocepahlus (AUC: 0.910-0.944), whereas DeepCNTD-Net achieved an AUC of 0.80-0.90 for edema and 0.90 for hydrocepahlus. These findings highlight the model’s reliability while pointing to areas for further refinement in neuro-trauma detection. Fig.~\ref{fig:det_examples} showcases multi-finding detection, illustrating precise trauma identification alongside occasional false positives due to confounding pathologies.

\begin{table}
\centering
\footnotesize 
\caption{Individual detection performance of DeepCNTD-Net with hemSegFeat and brainAnatFeat on the CQ500 dataset for hemorrhage, mass effect and midline shift.}\label{tab:det_perf_CQ500}
\begin{tabular}{|l|c|}
\hline
Finding &  AUC (95\% CI)\\
\hline
Hemorrhage & 0.920 (0.876-0.956) \\
Mass Effect & 0.840 (0.755-0.910) \\
Midline Shift & 0.965 (0.922-0.996) \\
\hline
\end{tabular}
\end{table}

\section{Conclusion}
This study highlights the potential of a specialized 3D foundation model for head trauma triage, addressing the need for rapid, accurate diagnostics in emergency medicine. By integrating LLM-driven automated labeling with task-specific neural networks for hemorrhage segmentation and brain anatomy parcellation, we developed a robust framework for detecting a broad spectrum of neuro-trauma findings in CT scans. Our DeepCNTD-Net variant achieved high accuracy across both common and less frequent critical conditions, with multimodal feature integration enhancing performance. It reached an average AUC of 0.861 for 16 neuro-trauma findings, reinforcing the importance of domain-specific pretraining. This work contributes to the growing evidence supporting AI-driven foundation models in clinical practice, helping bridge the gap between increasing emergency head CT scans and the radiologist shortage. It also serves as a research reference point for future foundation AI applications in neuro-trauma triage. To maximize clinical impact, future efforts will focus on real-world validation and seamless clinical integration to improve patient care.

\begin{credits}

\subsubsection{Disclaimer}
The concepts and information presented in this paper are based on research results that are not commercially available. Future availability cannot be guaranteed.
\end{credits}

%
%
%
\bibliographystyle{splncs04}

%





\end{document}